\documentclass[]{aa}
\usepackage{graphicx}
\usepackage{balance}
\usepackage[varg]{txfonts}
\usepackage{natbib}
\bibpunct{(}{)}{;}{a}{}{,} % to follow the A&A style
\usepackage[figuresright]{rotating}
\usepackage[a4paper,breaklinks]{hyperref}

\def\teff{$T\rm_{eff}$ }
\def\kms{$\mathrm{km\, s^{-1}}$}

\newcommand{\loggf}{\ensuremath{\log\,gf}}
\newcommand{\logg}{\ensuremath{\log g}}

\newcommand{\draftflag}{false}

\newcommand{\beq}{\begin{equation}}
\newcommand{\eeq}{\end{equation}}

\newcommand{\COBOLD}{{\sf CO$^5$BOLD}}

\newcommand{\cobold}{\COBOLD}

\newcommand{\vsini}{\ensuremath{V\sin(i)}}

\idline{12}{44}
\begin{document}

\title{GIANO Y-band spectroscopy  of dwarf stars: Phosphorus, Sulphur, and Strontium abundances
\thanks{Based on observations obtained with GIANO}}

\author{
E.~Caffau \inst{1,2},
S.~Andrievsky \inst{3,1},
S.~Korotin \inst{3},
L.~Origlia \inst{4},
E.~Oliva   \inst{5},
N.~Sanna   \inst{5}, 
H.-G.~Ludwig \inst{2},
P.~Bonifacio \inst{1}
}

\institute{ 
GEPI, Observatoire de Paris, PSL Resarch University, CNRS,
Universit\'e Paris Diderot, Sorbonne Paris Cit\'e, Place Jules Janssen, 92195
Meudon, France
\and
Zentrum f\"ur Astronomie der Universit\"at Heidelberg, Landessternwarte, K\"onigstuhl 12, 69117 Heidelberg, Germany
\and
Department of Astronomy and Astronomical Observatory, Odessa National University, Isaac Newton Institute of Chile, Odessa
Branch, Shevchenko Park, 65014, Odessa, Ukrain
\and
INAF, Osservatorio Astronomico di Bologna, Viale Ranzani 1, 40127 Bologna, Italy
\and
INAF, Osservatorio Astrofisico di Arcetri, Largo E. Fermi 5, 50125, Firenze, Italy
}
\authorrunning{Caffau et al.}
\titlerunning{Phosphorus, Sulphur, and Strontium abundances from GIANO spectra}
\offprints{E.~Caffau}
\date{Received ...; Accepted ...}

\abstract%
%\context
{In recent years a number of poorly studied chemical elements, such as phosphorus, sulphur, and strontium, have 
received special attention as important tracers of the Galactic chemical evolution.
}
%\aims
{By exploiting the capabilities of the infrared echelle spectrograph GIANO mounted at the Telescopio Nazionale Galileo, 
we acquired high resolution spectra of four Galactic dwarf stars spanning the metallicity range between about 
one-third and twice the solar value.
We performed a detailed feasibility study about the effectiveness of the  
P, S, and Sr line diagnostics in the Y band between 1.03 and 1.10\,$\mu$m.
}
%\method 
{
Accurate chemical abundances 
have been derived   using  one-dimensional model atmospheres
computed in local thermodynamic equilibrium (LTE).
We computed the line formation   assuming LTE for P, 
while we performed non-LTE analysis  to derive S and Sr abundances.
 }
%\results
{We were able to derive phosphorus abundance for three stars and an upper limit for one star, 
while  we  obtained the abundance of sulphur and strontium for all of the stars.
We find [P/Fe] and [S/Fe] abundance ratios consistent with solar-scaled or slightly depleted values,
while the [Sr/Fe] abundance ratios are more scattered (by $\pm$0.2 dex) around the solar-scaled value. This is
fully consistent with previous studies using both optical and infrared spectroscopy.
}
%\conclusions
{We verified that high-resolution, Y-band spectroscopy as provided by GIANO is a powerful tool to 
study the chemical evolution of P, S, and Sr in dwarf stars.
}
\maketitle

%%%%%%%%%%%%INTRODUCTION%%%%%%%%%%%%%%%%%%%%%%%%%%%%%%%%%%%

\section{Introduction}

%Giano
The infrared (IR) spectrograph GIANO \citep{origlia12} is mounted at the Nasmyth focus of the 
4\,m Telescopio Nazionale Galileo (TNG) in La Palma. 
It observes in the range 950-2450\,nm at a resolving power of 50000.
We exploit the capabilities of this new facility to perform a feasibility study aimed at verifying 
the effectiveness of high-resolution spectroscopy in the Y band 
to derive reliable abundances of phosphorus and two other elements, namely S and Sr, in Galactic dwarf stars.

%Phosphorus
Phosphorus, with an atomic number of 15, is a light element, which is abundant in the Universe
and essential for the life as we know it on Earth.
The systematic analysis of phosphorus in Galactic stars is somewhat recent
(see \citealt{pcrires11,roederer14}).
In fact, the presence of P in the stellar atmospheres of F-G stars can be 
revealed by \ion{P}{i} near infrared (IR) lines at about 1050\,nm or by 
ultra-violet (UV) lines at about 213\,nm \citep{roederer14}, which have to be observed
from space.
An exhaustive review of the chemical evolution of phosphorus and of its investigations in peculiar
stars can be found in the recent paper of \citet{roederer14}.

%Sulphur
Sulphur is an $\alpha$-element that is effectively 
produced in massive stars at the final stage of their evolution 
(SNe of type II). 
Determination of its abundance in F-G-K stars relies on a limited number of \ion{S}{i} lines 
in the visual and near-IR spectral range.
The situation with the lines available for measurements in the 
spectra of metal-poor stars becomes more complicated. Only 
the strongest IR \ion{S}{i} lines of the first (at about 920\,nm) and third multiplets\footnote{
Here and elsewhere in the paper we adopt the multiplet numbering of \citet{Moore}} 
are observable 
in the spectra of stars with metallicities $[{\rm M/H}]<-1.5$.
The lines of the first multiplet are very strong and also detectable  in extremely
metal-poor stars \citep{spite11}, but they are often contaminated by telluric absorption.
The lines of the third multiplet at 1045\,nm are not so strong, but clear from
telluric absorption and also very useful for abundance determination \citep{nissen07,ecmult3}
 at metal-poor regimes.
As shown by \citet{korotin09}, the NLTE
effects have different influences on  different \ion{S}{i} lines.

As for phosphorus, the systematic analysis of the chemical evolution of sulphur
is recent, but in the latest ten years this research has grown.
For an updated vision on the status of S chemical investigations, see \citet{matrozis13}.

%Strontium
Strontium is an astrophysically interesting element, since 
its abundance is often used as a measurement of the efficiency of the 
slow  neutron capture process in the intermediate mass stars.
Together with Y and Zr, strontium belongs to the peak of light s-process elements.
In the visual part of the spectrum there are only a few lines of
\ion{Sr}{ii}. Among them there are two resonant lines at 407.7\,nm and
421.5\,nm, and a subordinate line at 416.1\,nm.

All the lines in the visual part are blended to some extent. 
For instance, the wing of the resonant line at 407.7\,nm is 
distorted by the \ion{La}{ii} line at 407.73\,nm, \ion{Cr}{ii} line at 407.75\,nm, and 
\ion{Dy}{ii} line at 407.79\,nm. Another resonant line at  421.5\,nm has an even 
more distorted profile due to  blending with a strong \ion{Fe}{i} line 
at 421.543\,nm and molecular CN band. The subordinate line at 
416.1\,nm is situated in the red wings of the two strong \ion{Fe}{i} 
416.149\,nm and \ion{Ti}{ii} 416.153\,nm lines, and, moreover, it is 
blended with molecular bands of CN and SiH. 

Another problem with using  resonant \ion{Sr}{ii} lines for the abundance
determination is their weak sensibility to the abundance change
in stars of the solar metallicity (they are strong).
In contrast,  IR \ion{Sr}{ii} lines 
are free of these problems, they are practically unblended. 
Nevertheless, as it was shown by \citet{andrievsky11} 
they are strongly affected by the NLTE effects. Depending upon the atmosphere parameters 
and metallicity, the NLTE corrections can achieve up to $-0.5$\,dex and more.

In summary, reliable abundances of Sr  
in the atmospheres of cool stars of different metallicities can be
derived from their near-IR lines with the help of the
NLTE analysis only  \citep{spite11,andrievsky11}. 

We  present the chemical analysis of phosphorus, sulphur, and strontium in four unevolved stars observed with GIANO.
Two of these stars also have  CRIRES spectra and we find a good agreement with the P abundance derived by \citet{pcrires11}.
For sulphur, we compared the S abundance we derived from the GIANO spectra to observations taken with SOPHIE and 
to previous analysis based on multiplets eight and six.
We find a good agreement with the SOPHIE spectra, but not always with previous analysis.
For the analysis on Sr, we compared the results based on the GIANO observations
with ultraviolet lines in SOPHIE spectra and find a good agreement.

%%%%%%%%%%%%%%%%%%%%%%%%%%%%%%%%%%%%%%%%%%%%%%%%%%%%%%%%%%%%%%%%%%
\section{Observed spectra}

The GIANO spectra of the four Galactic dwarfs were acquired  during two technical nights on
October 22 (HD\,1355 and HD\,22484) and 23 (HD\,146 and HD\,22484), 2013.
GIANO is interfaced to the telescope with a couple of fibers mounted on the same connector
at a fixed projected distance on sky of 3\,arcsec.
We observed the science targets by nodding on fiber,
i.e. target and sky were taken in pairs and alternatively acquired on fiber A and B, respectively,
for an optimal subtraction of the detector noise and background.
HD\,1461 was observed by setting two pairs of AB exposures for a total on-source integration time of 20\,min, 
while for the other three stars (brighter) only one pair of AB spectra for a total exposure time of 10\,min  have been acquired.
We also observed  the hot (O6.5V) dwarf star HIP\,029216 as a telluric standard.

From each pair of exposures, an (A-B) 2D-spectrum has been computed.
As a result of the image slicer, each 2D  frame contains four tracks per order (two per fiber).
To extract and wavelength-calibrate the echelle orders from the 2D GIANO spectra,
we used the ECHELLE package in IRAF and some new,
ad hoc scripts grouped in a package named GIANO\_TOOLS, which can be retrieved at the TNG
WEB page\footnote{http://www.tng.iac.es/instruments/giano/gia\-no\_tools\_v1.2.0.tar.gz}.
We used 2D spectra of a tungsten calibration lamp taken in the daytime  to map the geometry of the
four spectra in each order and for flat-field purposes.
A U-Ne lamp spectrum was used  
for wavelength calibration.
We extracted the positive (A) and negative (B) spectra of the target stars  and summed them
together to get a 1D wavelength-calibrated spectrum with the best possible signal-to-noise ratio (SNR).

For the scientific purpose of this paper aimed at deriving the abundances of 
phosphorus, sulphur, and strontium, we focused our analysis on the 
GIANO spectral orders between \#76 and \#70, covering the Y band between 1.00 and 1.10\,$\mu$m.
This band is rather clean from telluric contamination and contains a number of suitable transitions 
for the chemical species we are interested in.

In order to make a comparison of S and Sr abundances derived
from IR and optical spectra, we retrieved from the SOPHIE archive
(\url{http://atlas.obs-hp.fr/sophie/}) spectra observed at 
Observatoire de Haute Provence for HD\,13555 
and HD\,22484. For HD\,13555, we retrieved 30 spectra observed
in the high resolution (HR) mode of SOPHIE ($R \approx 76500$),
observed between November 2006 and September 2008. The spectra were
coadded providing ${\rm S/N} \approx 700$ at 550\,nm.
For HD\,2483, we only retrieved two HR SOPHIE spectra observed
on October 23 and 26 2013  with exposure times of 200\,s and 150\,s, respectively.
The S/N of the summed spectrum is about 300 at 550\,nm. 
SOPHIE \citep{Bouchy,Perru08,Perru11} is an echelle spectrograph fiber-fed from
the Cassegrain focus of the 1.93\,m telescope at Observatoire de Haute-Provence (OHP).
It can work at high efficiency (HE) or HR,
corresponding to a resolving power of about 40\,000 and 80\,000, respectively.
SOPHIE spectra have a wavelength range from 387.2\,nm to 694.3\,nm.
SOPHIE spectra are reduced by the Geneva pipeline, which also provides 
radial velocity. 

%%%%%%%%%%%%%%%%%%%%%%%%%%%%%%%%%%%%%%%%%%%%%%%%%%%%%%%%%%%%%%%%%%
\section{Chemical abundance analysis}

We analysed two \ion{P}{i} lines of Mult.\,1 at 1052\,nm and  1058\,nm,
the three \ion{S}{i} lines of Mult.\,3, located at 1045\,nm, and
three lines of \ion{Sr}{ii} located at 1003, 1032, and 1091\,nm.
The atomic data we used are shown in Table\,\ref{irpline}.

\begin{table}
\caption{Infrared lines analysed in this work.}
\label{irpline}
\begin{center}
\begin{tabular}{lclr}
\hline
\noalign{\smallskip}
 Element & $\lambda$ [nm] & E$_{\rm low}$ & \loggf  \\
 {[nm]}    & vacuum/air     & {[eV]}\\
\noalign{\smallskip}\hline
\noalign{\smallskip}
\ion{P}{i} & 1053.241/1052.952 & 6.95 & $+0.240$ \\
\ion{P}{i} & 1058.447/1058.157 & 6.99 & $+0.450$ \\
\noalign{\smallskip}\hline
\noalign{\smallskip}
\ion{S}{i} & 1045.8316/1045.5449 & 6.86 & $+0.250$ \\
\ion{S}{i} & 1045.9622/1045.6757 & 6.86 & $-0.447$ \\
\ion{S}{i} & 1046.2272/1045.9406 & 6.86 & $+0.030$ \\
\noalign{\smallskip}\hline
\noalign{\smallskip}
\ion{Sr}{ii} & 1003.9405/1003.6654 & 1.805& $-1.312$ \\
\ion{Sr}{ii} & 1033.0139/1032.7309 & 1.839& $-0.353$ \\
\ion{Sr}{ii} & 1091.7864/1091.4887 & 1.805& $-0.638$ \\
\noalign{\smallskip}
\hline
\end{tabular}
\end{center}
\end{table}

For our four stars, we adopted the stellar parameters (${\rm T}_{\rm eff}$/\logg\/[Fe/H]) listed in Table\,\ref{pp}
from \citet{Chen,takada02,gonzalez10}. 
For each star, we computed a 1D-LTE model atmosphere with the code ATLAS\,12
in its Linux version \citep{kurucz05,sbordone04,sbordone05}.

Our sample of stars includes four unevolved stars.
The star HD\,10453 is an astrometric binary, the companion
is 1.3\,mag fainter and has moved from a distance
of about 4" in 1820 to 50 mas in 2011. This should not affect the spectra. 
The broadening in the spectrum seems much larger than 
that expected from the resolving power of 50\,000 of the spectrograph.
By investigating the \ion{P}{i} and \ion{S}{i} lines, 
we presume the star has a rotational velocity of at least 15\kms.
According to \citet{ammler12}, the stellar rotation is definitely lower (\vsini $=7.4\pm 0.4$\kms).

We analysed the SOPHIE spectra of the stars HD\,22484 and HD\,13555, which we used to confirm the S and Sr abundances we derived from GIANO spectra.
For both stars, we found a very good agreement with previous analysis.
The derived stellar parameters  (\teff/\logg/[Fe/H], microturbulence) are 5933/3.97/$-0.17$, 1.32\,km/s for  HD\,22484 and 
6470/3.83/$-0.18$, 1.47\,km/s for HD\,13555. The difference in [Fe/H] of about 0.1\,dex with
respect to the previous analysis for the latter star can be explained
by the lower microturbulence of 1.47\,km/s we derived, to be compared to 2.4\,km/s by \citet{takada02}.
HD\,13555 shows  a rotational velocity of about 10\,km/s in its SOPHIE spectrum, which
is also confirmed by the GIANO spectrum. 

%%%%%%%%%%%%%%%%%%%%%%%
\subsection{Phosphorus}

We measured the equivalent width (EW) of the \ion{P}{i} lines, by using the 
{\tt iraf}\footnote{Image Reduction and Analysis Facility, 
written and supported by the IRAF programming group at 
the National Optical Astronomy Observatories (NOAO) in Tucson, Arizona.
http://iraf.noao.edu/} task {\tt splot}
with a Gaussian profile for the line profile fitting.
The two \ion{P}{i} lines  we could detect (at 1052 and 1058\,nm) 
are blended with a \ion{Ni}{i} and \ion{Si}{i} line, respectively. 
According to the strength of the lines and the stellar
${\rm V}\sin{i}$, we fitted the \ion{P}{i} line alone, or took into account
the close-by line at the same time, using the deblending option of {\tt splot}.
The two \ion{P}{i} lines are present in both order 72 and 73.
We analysed both spectra and took as EW the average value.
We used the code WIDTH \citep{kurucz93,kurucz05,castelli05} 
to derive the P abundance from the EW values.
The results are listed in Table\,\ref{pp}. 

\begin{figure}
\begin{center}
\resizebox{\hsize}{!}{\includegraphics[draft = \draftflag, clip=true]{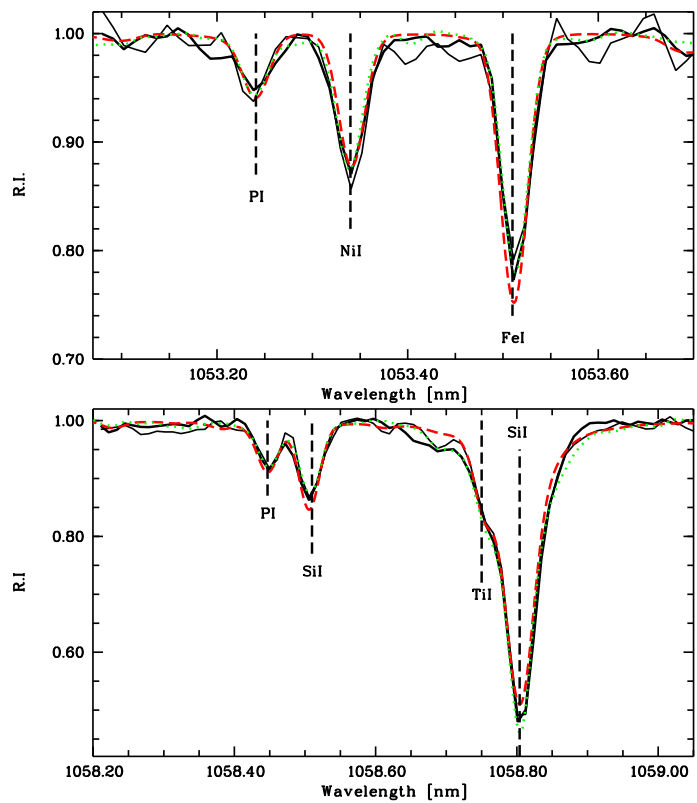}}
\end{center}
\caption[]{The two \ion{P}{i} lines are  shown (solid black line
thinner and thicker for order 72 and 73, respectively) in the case of HD\,1461, 
in comparison with a synthetic spectrum with A(P)=5.60 (dashed red).
The CRIRES spectrum is also over-plotted (dotted green) degraded at the 
resolution of GIANO.
}
\label{plothd1461}
\end{figure}

%HD1461
For HD\,1461, we have previous observations with CRIRES
and we find a good agreement between the abundances derived 
from the CRIRES and GIANO observations (see Fig.\,\ref{plothd1461}).

%%% FIGURE %%%%%%%%%%%%%%%%
%%% FIGURE %%%%%%%%%%%%%%%%

%HD13555
For star HD\,13555,
we did not use the order 72 to derive the EW of the line at 1053\,nm
because of the low SNR.
The agreement with the results from the CRIRES spectra is not perfect 
(about 0.1\,dex in P abundance) but is still good.

%HD10453
The only \ion{P}{i} line detectable in the spectrum of HD\,10453 is that at 1058\,nm,
but it is blended with a \ion{Si}{i} line and\ no P abundance can be derived because of the relatively
high rotation of the star. However, we can provide 
an upper-limit on A(P).

%HD22484
The spectrum of HD\,22484 is the best quality spectrum.
The two lines of \ion{P}{i} give abundances in 
good agreement, ${\rm A(P)}=5.36\pm 0.04$, but we decided to exclude  the line at 1052\,nm
from the order 72 owing to the low SNR.

\begin{table*}
\caption{\label{pp}
Stellar parameters and phosphorus abundances of our programme stars and comparisons,
when available, with the analysis on the CRIRES spectra from \citep{pcrires11}.
}
\begin{center}
\begin{tabular}{lllrllrrrrrrrr}
\hline\noalign{\smallskip}
  Target  & \teff & \logg & [Fe/H] & [S/H] & Ref  &  \multicolumn{2}{c}{EW [pm]} & 
\multicolumn{2}{c}{A(P)} & \multicolumn{2}{c}{EW [pm]} & \multicolumn{2}{c}{A(P)} \\
          &      &      &        &       &         &  Crires & Giano              & Crires & Giano & Crires & Giano & Crires & Giano \\
          &  K   &      &        &       &         &  \multicolumn{2}{c}{1053.2}&  \multicolumn{2}{c}{1053.2}  & \multicolumn{2}{c}{1058.4} & \multicolumn{2}{c}{1058.4}      \\
\hline\noalign{\smallskip}
 HD 1461   & 5765 & 4.38 & $+0.19$  &$-0.05$  & G10  & 2.00 & 2.10 & 5.59 & 5.62 & 2.70 & 2.70     & 5.59 & 5.59 \\    
 HD 10453  & 6368 & 3.96 & $-0.46$  &$-0.29$  & C02  &      &      &      &      &      & $<3.00$  &      & $<5.28$ \\  
 HD 13555  & 6470 & 3.90 & $-0.27$  &$-0.25$  & T02  & 1.80 & 2.34 & 5.14 & 5.28 & 2.90 & 3.13     & 5.22 & 5.27 \\     
 HD 22484  & 5960 & 4.02 & $-0.25$  &$-0.28$  & T02  &      & 1.90 &      & 5.33 &      & 3.00     &      & 5.41 \\     
\noalign{\smallskip}\hline\noalign{\smallskip}
\end{tabular}
\end{center}
The column ``Ref'' refers to the reference for the stellar parameters and
[S/H], and corresponds to 
G10: \citet{gonzalez10}; T02: \citet{takada02}; C02: \citet{Chen}.
\end{table*}

As the uncertainty in A(P) we quadratically added the observational uncertainty
(the scatter of the abundance derived from the two lines
added to the uncertainty in the EW measurement) to the systematic uncertainty 
related to uncertainty in the stellar parameters. 
We are aware of no NLTE study on phosphorus, nor of the existence of any model atom.
\citet{psun} investigated the granulation effects in the case of the Sun and derived very tiny effects
on these weak lines.

%%%%%%%%%%%%%%%%%%%%
\subsection{Sulphur}

\begin{figure*}
\begin{center}
\resizebox{\hsize}{!}{\includegraphics[draft = \draftflag, clip=true]{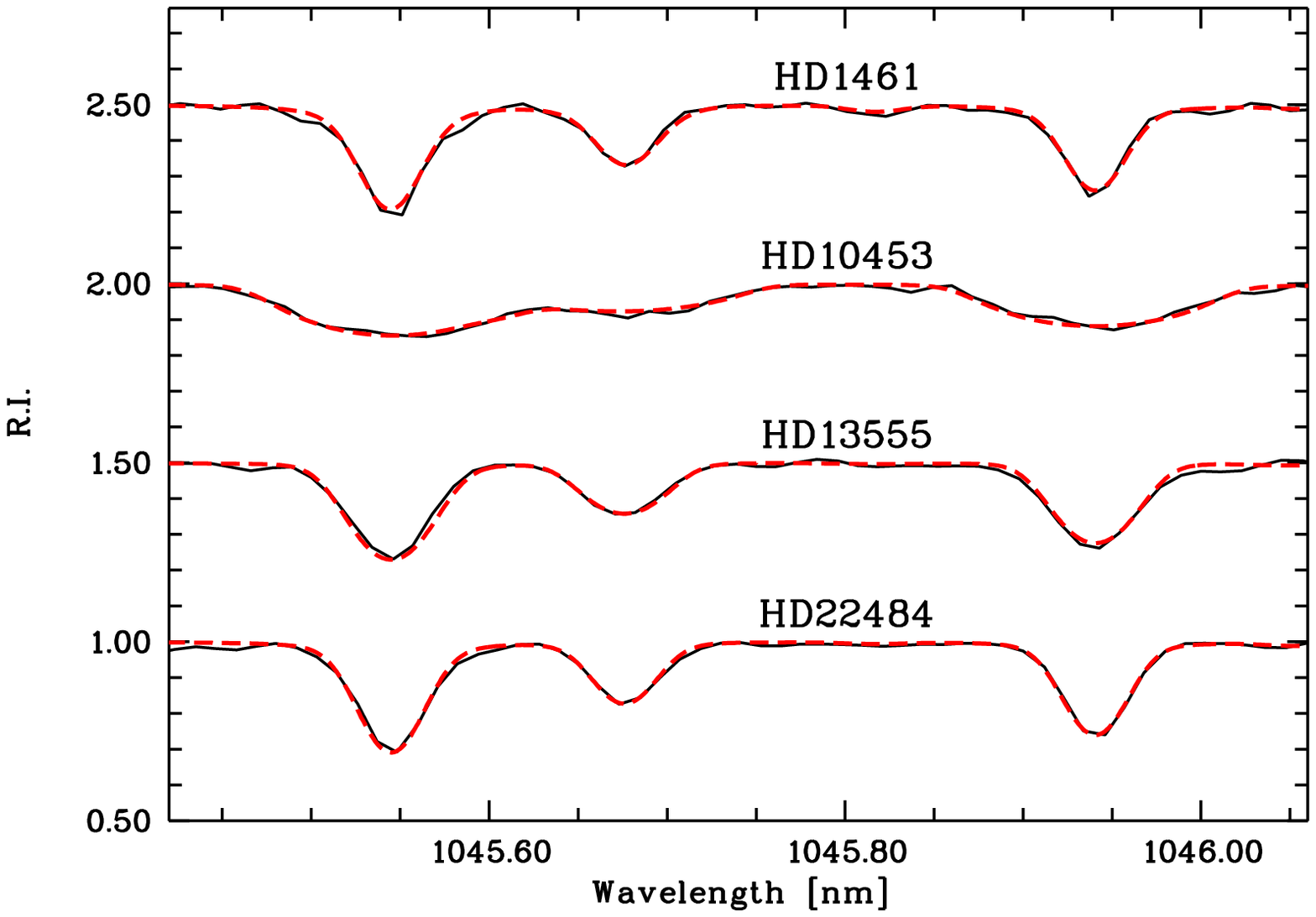}}
\end{center}
\caption[]{The three \ion{S}{i} lines of Mult.\,3 are  shown (solid black) 
in comparison with the best fit (dashed red).
For display purposes, the spectra are vertically displaced.
}
\label{pzolfi}
\end{figure*}

We derive the sulphur abundance by fitting the \ion{S}{i} lines of Mult.\,3 located at 1045\,nm
with NLTE profiles based on the ATLAS\,12 models.
The \ion{S}{i} lines of Mult.\,3 are clearly detected in all four stars
(see Fig.\,\ref{pzolfi})
and we derived the S abundances reported in Table\,\ref{pp}.
The three \ion{S}{i} lines have been fitted simultaneously so that we cannot
derive a line-to-line scatter. An uncertainty of 0.13\,dex takes 
the random (0.05\,dex) and systematic (0.12\,dex) uncertainties into account.

%%% FIGURE %%%%%%%%%%%%%%%%
%%% FIGURE %%%%%%%%%%%%%%%%

We compared a synthetic spectrum with the S abundance derived from the third multiplet
to the \ion{S}{i} lines of multiplet six in the SOPHIE spectra of each star and derived
a very good correspondence for all four stars.
We also find  a general good agreement in the A(S) with previous determination based on Mult.\,6 and 8 
for our stars except for the star HD\,13555.
For this star, the S abundance we derive from the third multiplet
in the GIANO spectra is about 0.3\,dex lower than the previous analysis by \citet{takada02} based on Mult.\,8.
We use the same stellar parameters, except for microturbulence, where they derive 2.4\,km/s
and we adopt 1.5\,km/s, which cannot explain the difference because a larger microturbulence would cause 
an even larger disagreement.

\citet{ecmult3} investigated the granulation effects on the A(S) determination from Mult.\,3, in the
solar case. The 3D corrections happen to be of the order of 0.1\,dex in the solar spectrum.
In the case of hotter stars, the 3D corrections are larger; e.g. for Procyon they became more than twice the solar case,
while at lower temperature they are smaller, e.g. for a 5000\,K dwarf star model they became negligible
(see \citealt{ecmult3} for details).

%%%%%%%%%%%%%%%%%%%%%
\subsection{Strontium}

\begin{figure}
\begin{center}
\resizebox{\hsize}{!}{\includegraphics[draft = \draftflag, clip=true]{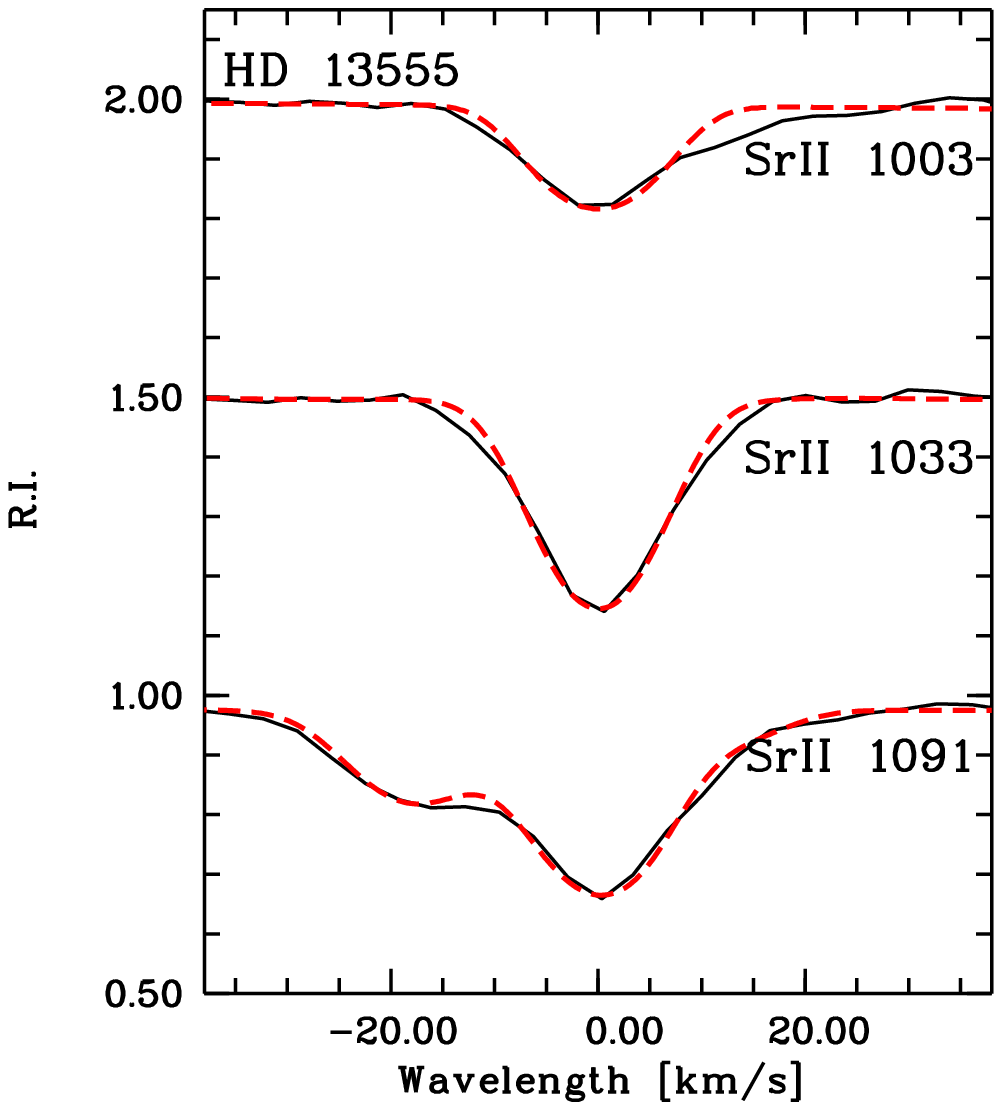}}
\end{center}
\caption[]{The three \ion{Sr}{ii} lines for HD\,13555 are  shown (solid black) 
in comparison with the best fit (dashed red).
For display purposes, the spectra are vertically displaced.
}
\label{psr13555}
\end{figure}

We derived the Sr abundances  by line profile fitting  using NLTE synthetic profiles,
and the Sr abundances we derived are presented in Table\,\ref{pp}.
In Fig.\,\ref{psr13555} we show the \ion{Sr}{ii} lines for HD\,13555.
The Sr abundances derived from the best fit on the \ion{Sr}{ii} lines in the GIANO spectra have
been used to synthesise profiles to compare to the \ion{Sr}{ii} lines in the
SOPHIE spectra, at 407.7, 416.1, 421.5, and 430.5\,nm.
The agreement is generally very good.
Two of our stars show a [Sr/Fe] of about +0.2 and one  a low value of $-0.1$.
This large scatter in [Sr/Fe] around solar metallicity is not unknown
and consistent with what is found  by \citet{mashonkina01}.

%%% FIGURE %%%%%%%%%%%%%%%%
%%% FIGURE %%%%%%%%%%%%%%%%

The uncertainty in the A(Sr) determination is of 0.12\,dex.
It is related to the uncertainty
in the line profile fitting (0.1\,dex, mainly due to continuum determination) and 
systematic uncertainty related to uncertainty in the stellar parameters (0.05\,dex).

\begin{table*}
\caption{\label{zolfo}
Stellar parameters, sulphur, and strontium abundances of our programme stars.
}
\begin{center}
\begin{tabular}{llllrrcccc}
\hline\noalign{\smallskip}
  Target  & \teff & \logg & [Fe/H]    & A(P)   & [P/Fe] &  {A(S)$_{\rm NLTE}$} & [S/Fe] & A(Sr) & [Sr/Fe] \\
          &  K   &      &           &        &        &  \\
\hline\noalign{\smallskip}                
 HD\,1461   & 5765 & 4.38 & $+0.19$ & 5.60    &$+0.05$ & 7.26 & $-0.09$ & 3.07 & $+0.04$ \\
 HD\,10453  & 6368 & 3.96 & $-0.46$ & $<5.28$ &$<+0.28$& 6.81 & $+0.06$ & 2.67 & $+0.21$ \\
 HD\,13555  & 6470 & 3.90 & $-0.27$ & 5.28    &$-0.09$ & 6.67 & $-0.22$ & 2.82 & $+0.17$ \\
 HD\,22484  & 5960 & 4.02 & $-0.10$ & 5.37    &$+0.01$ & 7.01 & $-0.05$ & 2.72 & $-0.10$ \\
\noalign{\smallskip}\hline\noalign{\smallskip}
\end{tabular}
\end{center}
\end{table*}

\begin{figure}
\begin{center}
\resizebox{\hsize}{!}{\includegraphics[draft = \draftflag, clip=true, angle=90]{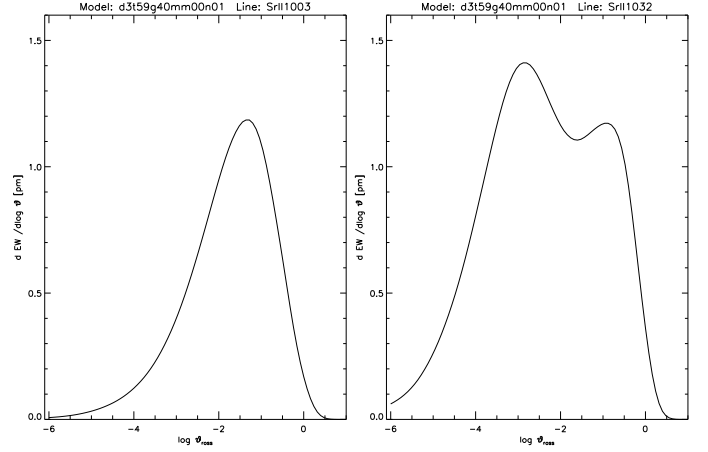}}
\end{center}
\caption[]{The contribution function for the EW for two \ion{Sr}{ii} lines in the case of the 
\cobold\ model with parameters 5900/4.0/0.0.
}
\label{plotsr}
\end{figure}

We investigated granulation effects for these \ion{Sr}{ii} lines by
analysing three hydrodynamical models (solar model, 5900/4.0/0.0, 6250/4.0/0.0) 
from the CIFIST grid \citep{ludwig09} 
computed with the code \COBOLD\ \citep{freytag12}.
The 3D effects are small for all three \ion{Sr}{ii} lines in the solar model,
lower than 0.08\,dex. 
These effects becomes larger (about 0.1\,dex for the line at 1003\,nm
and about 0.25\,dex for the other two lines) for a model with parameters 5900/4.0/0.0,
and a similar size for a hotter model with parameters 6250/4.0/0.0
(0.08\,dex for the line at 1003\,nm, 0.17\,dex for the line at 1032\,dex,
0.15\,dex for the line at 1091\,nm).
In Fig.\,\ref{plotsr}, the contribution function of the weakest (1003\,nm) and 
strongest (1032\,nm) \ion{Sr}{ii} lines, in the case of the 5900/4.0/0.0 model, is shown.
For the line at 1003\,nm, the contribution function has a smooth shape
with a maximum at about --1.5 in $\log\tau$.
In the case of the stronger 1032\,nm line, the contribution function shows
a double peak. This also happens in the case of the \ion{Li}{i}
doublet at 670.7\,nm as shown in Fig.\,1 by \citet{mst10}, where
the authors state that in the case of lithium that NLTE effects are large
and this is evident by comparing the contribution function for the 3D-LTE
and the 3D-NLTE synthesis.
We know that 1D-NLTE effects for the \ion{Sr}{ii} IR lines are somewhat large,
but we do not have 3D-NLTE computations at present.

%%% FIGURE %%%%%%%%%%%%%%%%
%%% FIGURE %%%%%%%%%%%%%%%%

%%%%%%%%%%%%%%%%%%%%%%%%%%%%%%%%%%%%%%%%%%%%%%%%%%%%%%%%%%%%%%%%%
\section{Discussion and conclusions}

\begin{figure*}
\begin{center}
\resizebox{\hsize}{!}{\includegraphics[draft = \draftflag, clip=true]{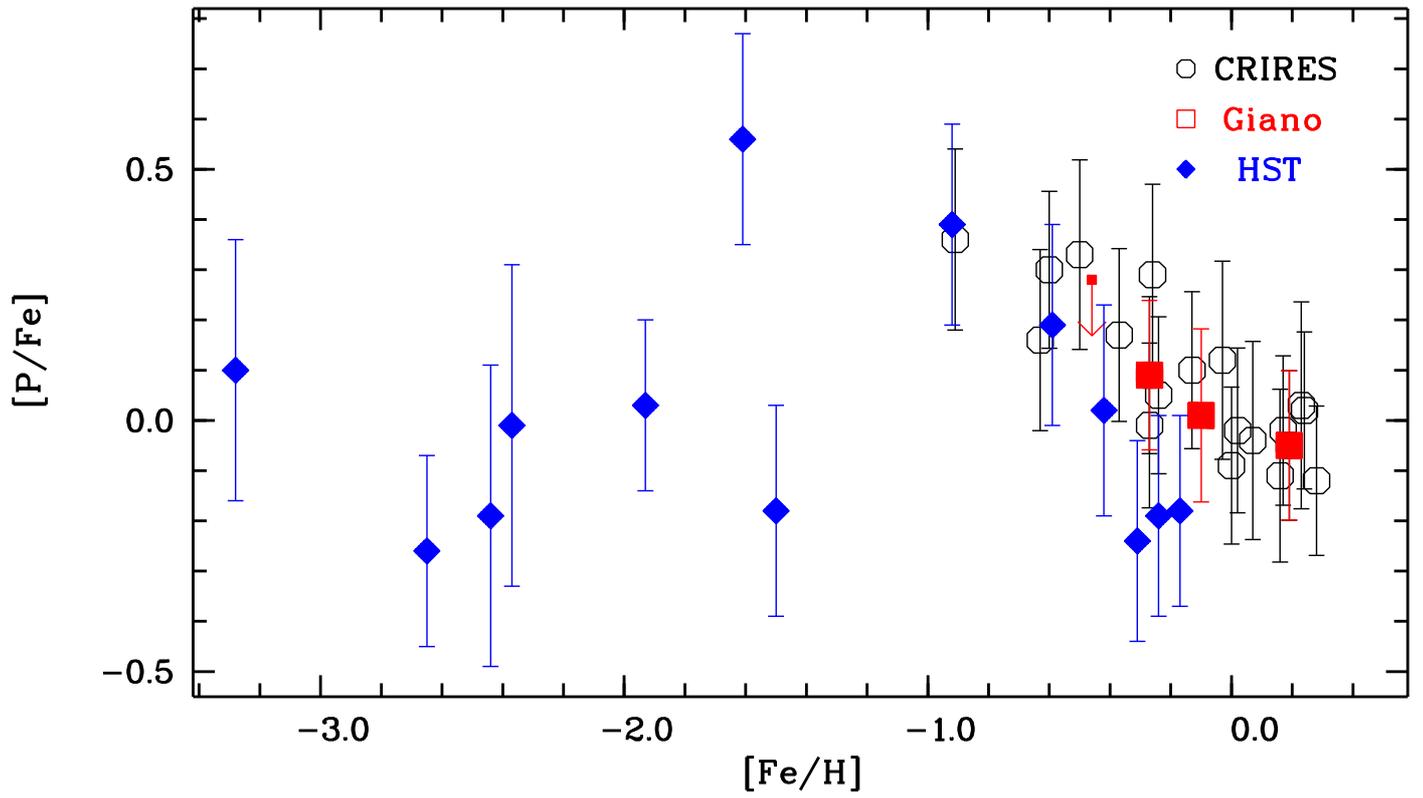}}
\end{center}
\caption[]{[P/Fe] vs. [Fe/H] is shown for the four stars  analysed
(solid red squares) in comparison with the results from Caffau et al. (2011; black open squares) 
and from Roederer et al. (2014; solid blue diamonds).
}
\label{fep}
\end{figure*}

In this work, we analysed the GIANO spectra in the Y band of four unevolved Galactic stars,
spanning a metallicity range between about one-third and twice the solar value,
with the purpose of measuring accurate abundances of P, S, and Sr.
For three out of four stars, we could derive P abundance, and for the fourth an upper limit.
For the two stars for which CRIRES spectra are available we find a concordance in the derived abundances.
Phosphorus abundances from the GIANO spectra also fit perfectly into the [P/Fe] versus [Fe/H], derived by  
\citet{pcrires11} and plotted in  Fig.\,\ref{fep}, 
where [P/Fe] smoothly decreases with increasing  stellar metallicity with solar-scaled values 
 within one $\sigma$ at [Fe/H]$\ge$0.0. 

The trend is similar to what is expected for an $\alpha$-element and, 
as discussed in \citet{cescutti12}, such a Galactic evolution of phosphorus
can be explained with P mainly produced in core-collapse supernovae with a minor
contribution from supernovae type Ia.
However, the yields have to be
increased by a factor of about three  to fit the observed abundances.
In Fig\,\ref{fep},  the abundances derived by \citet{roederer14} from UV lines
are also shown (blue solid diamonds) for comparison.
At its highest metallicities, these measurements show a similar behaviour in [P/Fe] vs. [Fe/H]
as ours, although the increase in [P/Fe] with decreasing [Fe/H] has a steeper slope.
At $\left[{\rm Fe/H}\right]<-1.5$, [P/Fe] show a constant value around the solar-scaled value.
\citet{jacobson14} explain this behaviour as a buildup of P with increasing Fe (see their Fig.\,2).
From the sample of \citet{roederer14} and according to \citet{jacobson14}, it is not so clear
which is the enhancing factor for the yields needed to reproduce the observed trend.

The GIANO Y-band \ion{S}{i} lines of the third multiplet are
particularly useful for deriving S abundance in metal-poor stars because 
they are strong, but not contaminated by strong telluric absorption as are the stronger
lines of the first multiplet. 
We derived S abundance from third multiplet for our stars with metallicities 
in the  $-0.46<\left[{\rm Fe/H}\right]<+0.19$ range,
and we find a good agreement when comparing synthetic profiles with the derived A(S) 
with weaker \ion{S}{i} lines, e.g. those of the sixth multiplet at optical wavelengths. 
We find [S/Fe] abundance ratios consistent with solar-scaled or slightly depleted values.

Three mostly clean \ion{Sr}{ii} lines are also present in the GIANO Y-band spectra of dwarf stars.
We derived Sr abundance from these IR lines and compared synthetic spectra computed
with our Sr abundances to optical \ion{Sr}{ii} lines in SOPHIE spectra, finding a
good agreement.
We find [Sr/Fe] abundance ratios scattered by $\pm$0.2 dex around the solar-scaled value.

This feasibility study has thus demonstrated that 
the GIANO spectra in the Y band are perfectly suited to derive P abundance in Galactic stars.
We proved that the A(P) derived from GIANO are consistent with those obtained with CRIRES, by using the same 
line diagnostics but higher spectral resolution. 
We also verified that S and Sr abundances as derived from the Y-band GIANO spectra are trustable and consistent with those obtained from 
different diagnostic lines at optical wavelengths.

%%% FIGURE %%%%%%%%%%%%%%%%
%%% FIGURE %%%%%%%%%%%%%%%%

%%%%%%%%%%%%%%%%%%%%%%%%%%%%%%%%%%%%%%%%%%%%%%%%%%%%%%%%%%%%%%%%%%%

\begin{acknowledgements}
The project was funded by FONDATION MERAC.
We acknowledge support from the Programme Nationale
de Physique Stellaire (PNPS) of the Institut Nationale de Sciences
de l'Universe of CNRS.
SMA and SAK acknowledge the SCOPES grant
No. IZ73Z0-152485 for financial support.
\end{acknowledgements}

\balance
\bibliographystyle{aa}

\end{document}